\documentclass[a4paper]{article}
\usepackage[ansinew]{inputenc}
\usepackage{times}
\usepackage[T1]{fontenc}
\usepackage{graphicx}
\usepackage{geometry}
\usepackage{amssymb}
\usepackage{amsmath}

\usepackage{rotating}

\usepackage{tcolorbox}
\usepackage{xcolor}
\colorlet{shadecolor}{gray!25}
\usepackage{subfig}
\usepackage{float}

\author{\normalsize Ludwig A. Hothorn,\\ 
\footnotesize Im Grund 12, D-31867 Lauenau, Germany (e-mail:ludwig@hothorn.de)\\ \scriptsize(retired from Leibniz University Hannover)}

\title{Seven recommendations for alternatives to the common analysis of variance (ANOVA)\\ with application in the life sciences - using R
}
\begin{document}

\maketitle
\begin{abstract}
Standard ANOVA is among the most widely used tests in the life sciences and beyond. Several alternatives are proposed to provide simultaneous confidence intervals, ensure tight control of FWER, be robust to variance heterogeneity, avoid pre-testing for global effect (for one-way designs) or irrelevant interaction (for multi-way designs) prior to multiple comparison procedures. On the basis of selected examples, the R programs are provided for this purpose. 
\end{abstract}

\section{Introduction}\label{sec1}
In bio/medical/life sciences, the analysis of variance (ANOVA) represents a widely used statistical method (cited over 170,000 times in the WebSci 04/2022)  for very different  issues and designs. Below, I focus on the considerable subset of low-dimensional, completely randomized designs (e.g., one- and two-way layouts) with a single primary endpoint and fixed effects error term. In the following, six recommendations are proposed, which lead to a more problem-adequate interpretation, adequate error control, and robustness to selected real data situations-  more or less a new kind of ANOVA: i) the use of  multiple comparisons (MCP) directly instead as post-hoc test after pre-test ANOVA, ii) the use of a cell means model for the primary factor instead of a pre-test on negligible interactions ????, iii) the use of analysis of means (ANOM) instead of ANOVA to provide simultaneous confidence intervals and extensions in the generalized linear model (GLM), iv) the use of a multiplicity-adjusted version instead of the common ANOVA, v) the appropriate analysis of treatment and time effects, iv) the use of  robust versions against heteroscedasticity, and vii) modeling dose (or concentration) both qualitatively (as MCP) and quantitatively (as regression model). \\
I start with a motivating example where using CRAN R packages for reproducible evaluation, see the related R-code.


\section{A motivating example}\label{sec2}
\begin{figure}[htbp]
	\centering
		\includegraphics[width=0.7\textwidth]{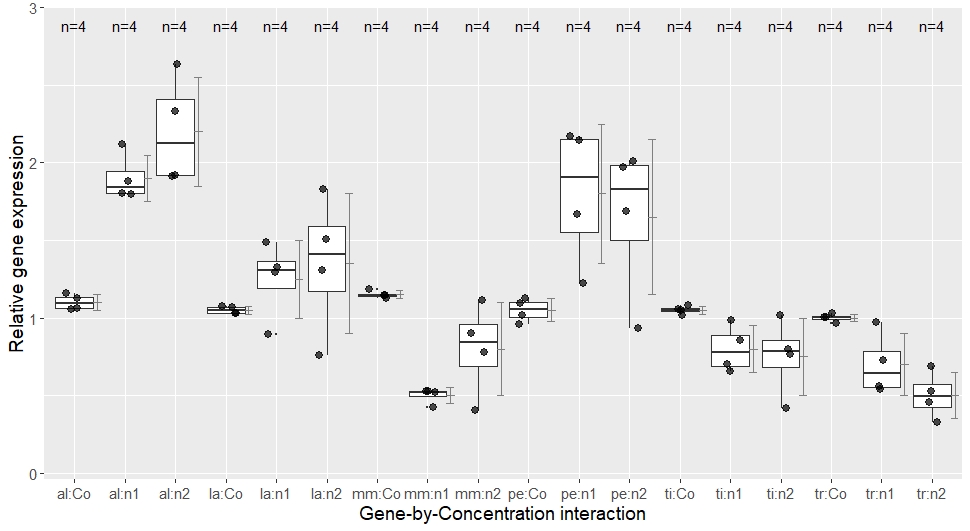}
	\caption{Boxplot for relative gene expression of six genes in MRC-5 cells were treated with TGF-beta}
	\label{fig:IAplot}
\end{figure}

Modeling asthma and fibrosis in-vitro, MRC-5 cells were treated with 0, 1 and 5 ng/ml TGF-beta (abbreviate with Co, n1, n2)  where the relative gene expressions were measured for six genes (alpha-SMA (al), Col1A1 (la), matrix metalloproteinase 1 (mm) periostin (pe), matrix metalloprotease 9 (ti) and  transient receptor potential ankyrin 1 (tr)) using qRT-PCR \cite{yap2021}.  Figure \ref{fig:IAplot} demonstrate a two-way layout with the factors genes and concentration (as a cell means pseudo one-way layout) in a balanced, small sample size ($n_i=4$) design using simulated data from the summary values assuming Gaussian distribution \cite{hkh2019} with obvious variance heterogeneity.

The investigators concluded an up-regulation of alpha-SMA(al) as well as of Col1A1 (la) and periostin (pe) as primary results. I take from this interpretation an one-sided question for increases, further as primary factor the genes and secondary factor the concentration, whereby already actually the gene-by-concentration interaction is of interest. The authors used two-way ANOVA followed by Bonferroni-adjusted two-tailed t-test between control and the two concentrations. With the help of this example I will demonstrate the above 6 recommendations of an alternative evaluation.

A major limitation of all these proposals is for designs with very small sample sizes, where $n_i=3$ is not at all exceptional in some bio-science studies. Not even reliable two-sample test for heterogeneity of variance or nonparametric location tests are available. Therefore, not only for reasons of test power, I recommend a minimum sample size of $n_i=5$.

\section{Recommendation I: Use analysis of means (ANOM) instead of ANOVA to provide simultaneous confidence intervals}\label{sec5}
A major disadvantage of the ANOVA F-test is that it is a global test only, but one is often interested in the magnitude of individual group or treatment differences.  This can be achieved, for example, by using subset F-test in a closure test \cite{LH2021y}. However, even here is the estimation of simultaneous confidence intervals difficult. While the ANOVA F-test is based on the sum of squared deviations of the group means from the overall mean, ANOM uses the maximum of their linear deviations and thus represents a MCP itself \cite{Pallmann2016}, whereby ANOM has a comparable power behavior as ANOVA \cite{Konietschke2013} under selected alternative hypotheses. I.e., one can replace the global F-test by the global ANOM test, as ANOM additionally offers either adjusted p-values or compatible simultaneous confidence intervals for the individual group comparisons due to the monotonicity properties of MCP's \cite{Gabriel1969}. In addition, ANOM allows directional testing (i.e. one-sided tests) and is available for generalized linear models (GLM), i.e. proportions or counts can also be analyzed adequately - considerable advantages. Furthermore, a related nonparametric version is available. Of course, for pure comparisons to the control, comparison to the overall mean is not very useful, so here we demonstrate comparison across all genes and concentrations in the cell means model:

\par\medskip 
\begin{tcolorbox}[width=15cm]
\sffamily \scriptsize
\begin{verbatim}
library(multcomp)
library(sandwich)
iatc$IA<-iatc$treat:iatc$con
mod1<-lm(mrc5~IA, data=iatc)                                # linear model for interaction factor
anova(mod1)                                                 # F-test standard ANOVA										
M1<-glht(mod1,linfct = mcp(IA = "GrandMean"), 
                       alternative="greater",vcov=vcovHC)   # MCP total mean
plot(M1)
\end{verbatim}
\rmfamily
\end{tcolorbox}
\par\medskip 

Not surprising is the p-value of the F-test tiny ($1.298e-14$) and but that of the MCT total means even smaller, but because of the approximation inaccuracies these are not comparable. In contrast, the simultaneous confidence intervals for the differences to the total mean clearly show significant increases for al(n1,n2) in Figure \ref{fig:CIplot}:

\begin{figure}[htbp]
	\centering
		\includegraphics[width=0.67891\textwidth]{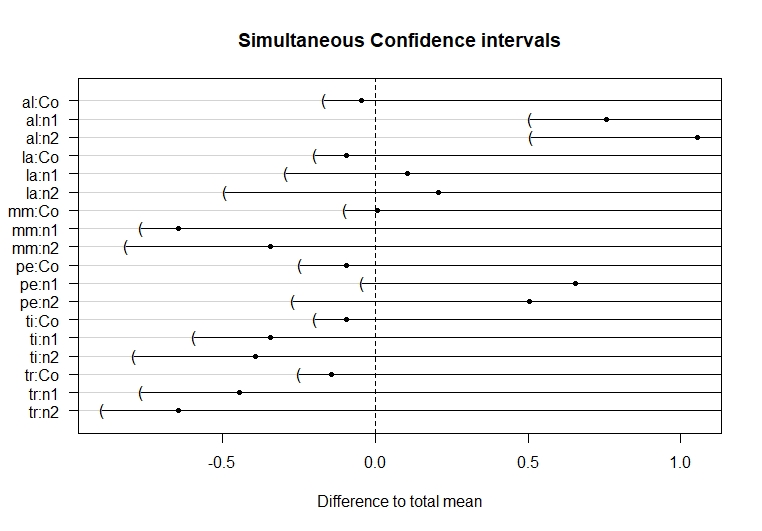}
	\caption{Confidence interval plot of the pseudo one-way cell means model for gene-by-concentration}
	\label{fig:CIplot}
\end{figure}

\subsection*{\small A nonparametric ANOM approach}\label{sec51}
If the normal distribution assumption for the primary endpoint 'relative gene expression' in the above example (see chapter \ref{sec3}) is too ambitious, one can alternatively use a nonparametric method, whereby the relative effect estimators can be recommended, especially if heterogeneous variances are possible \cite{Konietschke2012}. Based on the CRAN package \textit{nparcomp} \cite{ksh2015} the evaluation is similar to the parametric method above. 

\par\medskip 
\begin{tcolorbox}[width=15cm]
\sffamily \scriptsize
\begin{verbatim}
library(nparcomp)
nparcomp(mrc5~IA, data=iatc, asy.method = "mult.t",
         type = "Dunnett",alternative = "greater",
         plot.simci = FALSE, info = FALSE)
\end{verbatim}
\rmfamily
\end{tcolorbox}
\par\medskip 

However, because of the complete separation of the individual data in the treatment groups, invalid estimated test statistics and p-value result here.

\subsection*{\small ANOM approach in the GLM}\label{sec52}
Proportions are commonly summarized into 2-by-k contingency tables, e.g. the number of volunteers with nausea or without nausea after treatment with 0, 40, 80, 120 and 160 mg asnofaxine \cite{mi2022}:
\begin{table}[ht]
\centering\small
\begin{tabular}{rrrrrr}
  \hline
 & 0 & D40 & D80 & D120 & D160 \\ 
  \hline
Without nausea 		&  46 &  42 &  42 &  40 &  37 \\ 
With nausea 			&   3 &  10 &  10 &  11 &  14 \\ 
   \hline
\end{tabular}
\caption{2 by 5 contingency table for nausea after treatment with asnofaxine}
	\label{tab:tab}
\end{table}

Although a global test for such a safety parameter is not useful in a dose-finding study, it is performed here for methodological reasons. 

\par\medskip 
\begin{tcolorbox}[width=15cm]
\sffamily \scriptsize
\begin{verbatim}
library(multcomp)
mod12 <- glm(na ~ Dose, data=naus, family= binomial(link="identity"))
summary(glht(mod12, linfct = mcp(Dose = "GrandMean")))
chisq.test(table(naus))
\end{verbatim}
\rmfamily
\end{tcolorbox}
\par\medskip 

While the common-used Pearson $\chi^2$-test for 2 by k table data gives a p-value of 0.095, the ANOM-GLM for the difference between proportions shows a much smaller p-value of 0.002, even in its two-sided version. There are several test variants, e.g. odds ratio instead of risk difference as effect size.
 
\subsection*{\small An ANOM approach for ratios instead of differences}\label{sec53}
Although the difference in expected values is the widely used effect estimator, alternative ratios can be used. Due to the CRAN package \textit{mratios}, the estimation of adjusted p-values or simultaneous confidence intervals compatible to them is quite simple, using a Welch-type modification (simtest.ratioVH) here:
\par\medskip 
\begin{tcolorbox}[width=15cm]
\sffamily \scriptsize
\begin{verbatim}
library(mratios)
simtest.ratioVH(mrc5~IA, data=iatc,type = "GrandMean", alternative = "two.sided")
\end{verbatim}
\rmfamily
\end{tcolorbox}
\par\medskip 
In both approaches (differences, ratios), the combination 'mm:nn1' is furthest from the overall mean, the tiny p-values cannot be compared quantitatively due to the approximation limit.

\normalsize
\section{Recommendation II: Use multiple comparisons directly rather than a post-hoc test after pretest ANOVA}\label{sec3}
The very term 'post-hoc' for multiple comparison procedures associates that MCP conditionally (i.e., only if previous ANOVA is significant) (or unconditionally) follow an ANOVA pretest. Arguments against this are different alternative spaces of ANOVA and MCP (e.g. ANOVA and one-sided Dunnett \cite{Dunnett1955} tests \cite{LH2016}), a problematic error control of conditional tests, the differences in quality between pre- and post-hoc tests, the complexity of the procedure, problems to estimate adjusted p-values or even simultaneous confidence intervals, and others. For one-way designs, I recommend omitting inference using ANOVA F-test at all. I.e., ANOVA is used only to estimate expected means, means square error, and degree of freedom.  For two-way layouts (and higher one), I also recommend omitting inference using the F-tests of main and interaction effects, see Chapter \ref{sec4} for further detail. In the above example, let's assume that periostin (pe) would be interested in which concentration has increased relative expression compared to the control, then I would directly use the (pseudo)one-way Dunnett procedure:

\par\medskip 
\begin{tcolorbox}[width=15cm]
\sffamily \scriptsize
\begin{verbatim}
library(multcomp)
library(sandwich)
iatc3<-iatc[iatc$treat=="pe", ]               # select the data for pe
mod6<-lm(mrc5~conc, data=iatc3)               # linear model for one-way layout
anova(mod6)                                   # F-test for standard ANOVA
summary(glht(mod6,linfct = mcp(conc = "Dunnett"), 
        alternative="greater", vcov=vcovHC))  # one-sided Dunnett test using sandwich estimator

\end{verbatim}
\rmfamily
\end{tcolorbox}
\par\medskip 

The p-value of the standard ANOVA F-test of $p=0.054$ suggests a stop with the conclusion: no heterogeneity between these three groups, while the p-values of the Dunnett procedure (adjusted for possible variance heterogeneity, see chapter \ref{sec12} shows a clear increase at 1 ng (p=0.019). Why an ANOVA pretest for one-way designs at all?

\subsection*{\small Use non-parametric multiple comparisons directly}\label{sec31}
Based on the CRAN package nparcomp \cite{ksh2015} the evaluation is quite similar to the parametric method above (for the periostin (pe) subset):

\par\medskip 
\begin{tcolorbox}[width=15cm]
\sffamily \scriptsize
\begin{verbatim}
library(nparcomp)
nparcomp(mrc5~conc, data=iatc3, asy.method = "mult.t",
         type = "Dunnett",alternative = "greater",
         plot.simci = FALSE, info = FALSE)
\end{verbatim}
\rmfamily
\end{tcolorbox}
\par\medskip 
The pattern of p-values is similar to the above parametric approach, however the p-value for the comparison $n1- Co$ is much smaller $6.5e-11$.

\subsection*{\small Use multiple comparisons in the GLM directly}\label{sec32}
For safety data, the multiple comparison to the control is more problem-adapted than comparisons to the overall mean. In the original paper \cite{mi2022}, one-sided Fisher pairwise tests for increase are reported, actually non-inferiority tests would be indicated. Here, one-sided Dunnett-type comparisons are used for difference to control:

\par\medskip 
\begin{tcolorbox}[width=15cm]
\sffamily \scriptsize
\begin{verbatim}
library("multcomp")
nausCI <- summary(glht(mod12, linfct = mcp(Dose = "Dunnett"),alternative="greater"))
\end{verbatim}
\rmfamily
\end{tcolorbox}
\par\medskip 

\begin{table}[ht]
\centering\small
\begin{tabular}{l|rr}
  \hline
  Contrast & test stat. & p-value \\ 
  \hline
D40 - 0 & 2.03 & 0.077 \\ 
D80 - 0 & 2.03 & 0.077 \\ 
D120 - 0 & 2.31 & 0.040 \\ 
D160 - 0 & 2.99 & 0.006 \\ 
   \hline
\end{tabular}
\caption{Dunnett-type comparisons for the nausea incidences after treatment with asnofaxine}
	\label{tab:naus}
\end{table}

Depending on what one would tolerate in terms of increased nausea, doses below 120 are probably still acceptable (see the non-significant p-values in Table \ref{tab:naus}). 

\subsection*{\small Use multiple comparisons for ratio-to-control directly}\label{sec33}
A competitive procedure is Dunnett-type comparisons for ratios-to-control:

\par\medskip 
\begin{tcolorbox}[width=15cm]
\sffamily \scriptsize
\begin{verbatim}
library(mratios)
simtest.ratioVH(mrc5~conc, data=iatc3,type = "Dunnett", alternative = "greater")
\end{verbatim}
\rmfamily
\end{tcolorbox}
\par\medskip 

The both p-values are 0.043 for $n1/Co$ and 0.090 for $n2/Co$ compared with the p-values for differences-to-control of 0.019, 0.067. The choice between additive and multiplicative model depends however not on the magnitude of the p-value but on the effect interpretation.

\section{Recommendation III: In two-way layouts use cell means model for the primary factor instead of a pretest on negligible interactions}\label{sec4}

In two-way layouts, a non-significant interaction test is commonly used to make multiple comparisons between levels of the primary factor pooled over the levels of the secondary factor- or for significant interaction, sliced for each level of the secondary factor (each at level $\alpha$. This approach reveals several difficulties and inconsistencies, among them that the pretest on tolerable interaction should be formulated as a test on equivalence. An alternative is the simultaneous analysis of pooled and sliced data without any pretest 
\cite{hothorn2022x}. In the following, I consider \textit{conc} as primary factor and \textit{gene} as secondary factor in the subset data \textit{iatcE} (with genes al and la only for didactically purpose):

\par\medskip 
\begin{tcolorbox}[width=15cm]
\sffamily \scriptsize
\begin{verbatim}
Mod2<-lm(mrc5~conc+treat+conc:treat, data=iatcE)     # global ANOVA
f99<-anova(Mod2)					                            # F-test
#Mod1<-lm(mrc5~conc+treat, data=iatcE)                 # model without IA

iat1<-droplevels(iatcE[iatcE$treat=="la", ])          # sliced for la
iat2<-droplevels(iatcE[iatcE$treat=="al", ])          # sliced for al
#Mod3<-lm(mrc5~conc, data=iat1)                        # model for la
#Mod4<-lm(mrc5~conc, data=iat2)                        # model for al

Du <- contrMat(table(iatcE$conc), "Dunnett")
CM1 <- cbind(Du, matrix(0, nrow = nrow(Du), ncol = ncol(Du)))
rownames(CM1) <- paste(levels(iatcE$treat)[1], rownames(CM1), sep = ":")
CM2 <- cbind(matrix(0, nrow = nrow(Du), ncol = ncol(Du)), Du)
rownames(CM2) <- paste(levels(iatcE$treat)[2], rownames(CM2), sep = ":")
CM3 <- rbind(CM1, CM2)
colnames(CM3) <- c(colnames(Du), colnames(Du))           # partial Dunnett matrix 

iatcD$ww <- with(iatcE, interaction(conc,treat))       # interaction term 
#Mod5 <- lm(mrc5 ~ ww - 1, data = iatcE)                # cell means model
#X5 <- summary(glht(Mod5, linfct = CM3))                  # simult. cell means  

CM4<-rbind("pooled: 1 - 0" = c(-0.5, 0.5,0, -0.5, 0.5,0),
         "pooled: 5 - 0" = c(-0.5, 0, 0.5, -0.5,0, 0.5))
CM5<-rbind(CM1,CM2,CM4)                                    # joint matrix

library(sandwich)
summary(glht(Mod5, linfct = CM5, 
            alternative="greater", vcov = sandwich))   # simult. jointly 
\end{verbatim}
\rmfamily
\end{tcolorbox}
\par\medskip 
The unadjusted ANOVA F-tests reveals significant conc,  gene and interaction effects (but unknown which one). The joint analysis of sliced and pooled data reveals a strong increase of concentration 1 vs control and a also increase of concentration 5 vs. control in the pooled analysis, no increases in \textit{gene=la} only, whereas a remarkable increase in \textit{gene=al}. I.e. without any pretest, the significant increases vs. control in \textit{gene=al} can be demonstrated.

\begin{table}[ht]
\centering\small
\begin{tabular}{r|l|r}
  \hline
Data type & Comparison vs. control & adj. p-value\\ 
  \hline
Sliced for gene=al 		& 1 - 0 & 0.0000000002 \\ 
											& 5 - 0 & 0.0000016694 \\ 
Sliced for gene=la    & 1 - 0  & 0.1687254024  \\ 
                      & 5 - 0 & 0.2691304032 \\ 
Pooled                & 1 - 0 & 0.0000004012 \\ 
                      & 5 - 0 & 0.0000472096 \\ 
   \hline
\end{tabular}
\caption{Joint evaluation of a two-way design with possible interaction: adjusted p-values} 
\end{table}

\section{Recommendation IV: Use factorial ANOVA adjusted for multiplicity}\label{sec7}
In factorial ANOVA the multiple F-tests (on main factors, on interactions) are used commonly without control of the familywise error rate (FWER), i.e. for the $\xi$ F-tests $\rightarrow$ the $\xi$ marginal p-values are interpreted each at level $\alpha$. The main argument is the stochastic independence of the underlying sum-of-squares decomposition. There are only few paper discussing the incomplete FWER control  in factorial designs, e.g. \cite{cramer2016hidden} proposed simple  Bonferroni-adjustment (based on stochastic independence) and 
\cite{LH2022t} proposed multiple ANOM-tests, i.e. $\zeta$-variate t-distributed maxT-tests. Although the per-factor tests are uncorrelated, the remaining correlation matrix in the general maxT approach  is complex. Such a FWER control is conservative. In the subset \textit{iatcC} (using genes la, pe,ti only for didactically reasons) the above two-way layout example is formulated as:

\par\medskip 
\begin{tcolorbox}[width=15cm]
\sffamily \scriptsize
\begin{verbatim}
mod91<-lm(mrc5~conc+treat+conc:treat, data=iatcC)     # ANOVA factor model
f91<-anova(mod91)                                     # ANOVA F-tests
pun<-f91[1:3,5]                                       # unadj. F-tests
pad<-pun*3                                            # Bonfer.adj. F-tests
library(multcomp); library(emmeans); library(sandwich)
rg<-ref_grid(mod91, vcov. = sandwich::vcovHAC)        # sandwich estimator
nn <- emmeans(rg, ~ conc * treat)                     # interaction model
f1 = contrast(emmeans(nn, "conc"), "eff")             # vs. overall mean contrasts for conc
f2 = contrast(emmeans(nn, "treat"), "eff")            # vs. overall mean contrasts for treat
f1f2 = contrast(nn, "eff")                            # vs. overall mean contrasts for 
																											# conc-by-gene interaction
all =rbind(f1,f2,f1f2)                                # combination of the 3 contrast objects
mANOM<-summary(as.glht(all))                          # multiplicity adjustment (multi-t)
\end{verbatim}
\rmfamily
\end{tcolorbox}
\par\medskip 

\begin{table}[ht]
\centering\footnotesize
\begin{tabular}{ll|rrrr}
  \hline
Factor & Levels & estimate & se & t & p-value \\ 
  \hline
Conc & 0-GM & -0.141 & 0.038 & -3.82 & 0.00689 \\ 
			& 1-GM & 0.089 & 0.035 & 2.58 & 0.11632 \\ 
				& 5-GM & 0.056 & 0.063 & 0.88 & 0.94964 \\ 
 Gene  & la-GM & 0.022 & 0.051 & 0.43 & 0.99894 \\ 
				& pe-GM & 0.306 & 0.036 & 8.59 & 1.9e-08 \\ 
				& ti-GM & -0.328 & 0.045 & -7.35 & 3.4e-07 \\ 
  Interactions  & 0 by la & -0.144 & 0.038 & -3.78 & 0.00744 \\ 
 							& 1 by la & 0.055 & 0.076 & 0.73 & 0.97925 \\ 
  							& 5 by la & 0.156 & 0.193 & 0.81 & 0.96658 \\ 
  							& 0 by pe & -0.144 & 0.039 & -3.73 & 0.00863 \\ 
 							& 1 by pe & 0.606 & 0.059 & 10.20 & 2.5e-10 \\ 
							& 5 by pe & 0.456 & 0.108 & 4.20 & 0.00252 \\ 
							& 0 by ti & -0.144 & 0.039 & -3.69 & 0.00963 \\ 
							&  1 by ti & -0.394 & 0.055 & -7.22 & 8.4e-07 \\ 
							& 5 by ti & -0.444 & 0.121 & -3.69 & 0.00907 \\  \hline\hline
	ANOVA unadj & Conc &  & & & 0.139 \\ 
							& Gene &  & & & 8.9e-05  \\ 					
							& Conc-by-Gene &  & & & 0.0192 \\ 		\hline
	ANOVA Bonf. adj & Conc &  & & & 0.416 \\ 
							& Gene &  & & & 0.00027 \\ 					
							& Conc-by-Gene &  & & & 0.058 \\ 		
   \hline
\end{tabular}
\caption{ANOM Adjusted} 
\label{tab:ex14}
\end{table}

\section{Recommendation V: Evaluate repeated measures in terms of inference appropriately}\label{sec11}
There are many different motivations for using repeated measures, and the evaluation strategies and models are correspondingly diverse. The dependence of the $t^{th}$ observation on the $(t-1)^{th}$ observation within a subject must always be modeled correctly, e.g. via a suitable correlation structure in the mixed effect model. Somewhat lacking in these many models is the inference of treatment \textbf{and} time effects, e.g., which dose shows the highest effect at which time point. This complex correlation structure of unrelated treatment and related time contrasts can be estimated relatively easily using time-point-specific multiple marginal models \cite{Pipper2012}, with several possibilities of simultaneous inference available in the \textit{SimLongi} package \cite{Pallmann2018}. As an example, the heart rates of two drugs (ax23, bww9) in comparison with control were measured for  the four time points $T1, T2, T3, T4$ on the same subject \cite{Pallmann2017}. 

\begin{figure}[htbp]
	\centering
		\includegraphics[width=0.790\textwidth]{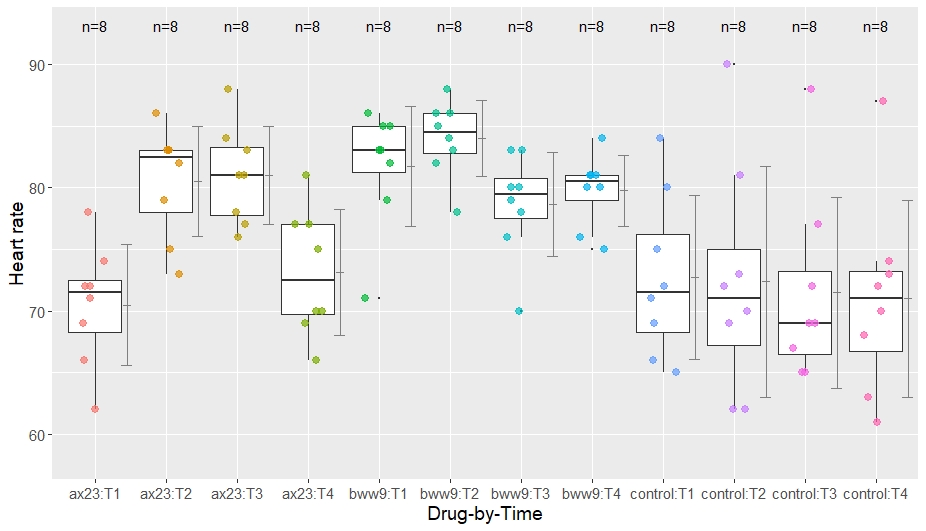}
	\caption{Treatment-by-time relationship in the heart rates example}
	\label{fig:Timeplot}
\end{figure}

The R-code is formulated for Dunnett-type between-group comparisons and between-time comparisons, jointly:

\par\medskip 
\begin{tcolorbox}[width=15cm]
\sffamily \scriptsize
\begin{verbatim}
library(SimLongi)
data(heart)
SMM<-SimLongiMMM(data=heart, response="heartrate", group="drug",
                 time="time", id="person", type="Dunnett",
                 alternative="greater", base=3, refdist="t")$Results
\end{verbatim}
\rmfamily
\end{tcolorbox}
\par\medskip 

Table \ref{tab:tt} reveals that the highest heart rates are achieved for drug ax23 at time T3 and for drug bww9 at time T2, with bww9 showing a significant increase already after T1.

\begin{table}[ht]
\centering\small
\begin{tabular}{ll|rrrrrr}
  \hline
Time-point & Treatment contrast & estimate & se & test stat & p-value\\ 
  \hline
T1 & ax23 - control & -2.25 & 2.76 & -0.81 & 0.969 \\ 
 &bww9 - control & 9.00 & 2.97 &3.03 & 0.014 \\ 
T2 &ax23 - control & 8.13 & 3.13  & 2.59 & 0.035 \\ 
  & bww9 - control & 11.63 & 3.560 & 3.27 & 0.008 \\ 
  T3 &ax23 - control & 9.50 & 2.79  & 3.40 & 0.006 \\ 
  &bww9 - control & 7.13 & 3.27  & 2.18 & 0.078 \\ 
  T4 &ax23 - control & 2.13 & 2.86  & 0.74 & 0.536 \\ 
  &bww9 - control & 8.75 & 2.95  & 2.97 & 0.016 \\ 
   \hline
\end{tabular}
\caption{Dunnett-type comparisons for treatment-by-time}
	\label{tab:tt}

\end{table}

\section{Recommendation VI: Use robust versions against heteroscedasticity}\label{sec12}
A source of bias that should not be underestimated is a pattern of group-specific heterogeneous variances, especially in the unbalanced design. Now, there are some methods for heteroscedasticity in the linear model, but here it is exactly about group-specific variance increases/decreases with inverse numbers of cases (small $n_i$, large $n_i$). The usual MQR estimator uses a kind of weighted average over the variances of all groups and favors exactly those groups with $ \Uparrow s_i^2/\Downarrow n_i$ and disfavors groups with $ \Downarrow s_i^2/\Uparrow n_i$ on the other hand. Then one should use k-sample Welch-type modifications \cite{Hasler2008} or simply instead of the MQR estimator the sandwich estimator \cite{Zeileis2006}, which is a suitable modification for MCP with not too small sample sizes \cite{Herberich2010} and can be easily realized with the CRAN package \textit{sandwich}. Since a conditional pretest for heterogeneity is not useful \cite{Kluxen2020}, this modification can be recommended as a standard method. The tiny bias can be easily seen for the subset pe-gene (see Chapter \ref{sec3}) in the comparisons versus control in this balanced design:

\begin{table}[ht]
\centering
\begin{tabular}{l|r|r|r}
  \hline
  Comparison &  p-value sandwich estimator & p-value MQR estimator \\ 
  \hline
1 - Co  & 0.019 & 0.021 \\ 
5 - Co  & 0.067 & 0.051 \\ 
   \hline
\end{tabular}
\end{table}

\newpage


\section{Recommendation VII: Is the use of ANOVA-type methods appropriate to analyse quantitative covariates?}\label{sec13}
One can simply convert quantitative covariates, such as doses or concentrations, into a qualitative factor and then analyze them using ANOVA-type methods (including MCP). A loss of power may occur and therefore it is warned against it \cite{greenland1995}. This is not quite so clear-cut, as the doses do not have to be proportional to the concentrations at the target, and higher doses could plateau. I.e. sometimes the joint testing of dose as a quantitative covariate, e.g. as Tukey-trend test \cite{Tukey1985} using ordinal, arithmetic and logarithmic dose scores  \cite{Schaarschmidt2021} and qualitative factor levels, e.g. as Williams-trend test \cite{Williams1971} represents an alternative \cite{HRSetal}. For the above  0, 1 and 5 ng/ml TGF-beta concentrations in the gene alpha-SMA (al)

\par\medskip 
\begin{tcolorbox}[width=15cm]
\sffamily \scriptsize
\begin{verbatim}
library(tukeytrend)
iatcA<-iatc[iatc$treat=="al", ]                            # data subset gene=al
mod81<-lm(mrc5~conc, data=iatcA)                           # ANOVA model conc... factor
anova(mod81)                                               # ANOVA F-test
mod82<-lm(mrc5~Conc, data=iatcA)                           # linear model Conc ... covariate
test82 <- tukeytrendfit(mod82, dose="Conc", ddf="residual",
          scaling=c("ari", "ord", "arilog", "treat"), ctype="Williams")
					                                                   # joint regression and Williams-MCP
JointTukeyWilliams<-summary(glht(test82$mmm, test82$mlf,
											vcov = sandwich, alternative="greater"))  # one-sided tests, sandwich variance est.
\end{verbatim}
\rmfamily
\end{tcolorbox}
\par\medskip

\begin{table}[ht]
\centering \footnotesize
\begin{tabular}{ll|rrr|r}
  \hline
Method& Model & estimate & se & t & p-value \\ 
  \hline
Tukey trend & arith conc. score       & 0.18 & 0.04 & 4.35 &  1.19e-05 \\ 
            & ordinal conc. score     & 0.55 & 0.08 & 6.71 & 1.05e-11 \\ 
            & (arith) logarithm. conc. score & 0.44 & 0.10 & 4.35& 1.14e-05 \\ 
Williams test & $5 - 0$ & 1.10 & 0.15 & 7.19 & 3.93e-13 \\ 
              & $(5+1)/2-0$ & 0.95 & 0.08 & 11.1 & $<1.0e-13$ \\ \hline \hline
ANOVA         & F-test &  &  &  & 0.000174 \\ 							
   \hline
\end{tabular}
\caption{Adjusted p-values for Tukey-Williams trend test and ANOVA F-test} 
\label{tab:ex82}
\end{table}

Although all tiny p-values are clearly in $H_1$, the F-test is less sensitive, whereas the linear regression reveals a smaller p-value, but the smallest p-value occurs for the plateau Williams-contrast of both concentrations in this example. A didactically suitable example of the interplay between quantitative and qualitative modeling of the concentration.

\normalsize

\section{Summary}\label{sec14}
\normalsize
If some of these seven recommendations are followed, the evaluation of factorial designs can be more problem-adequate.
Therefore, the impressive simplicity of ANOVA is lost, but the interpretability, especially of confidence intervals, should be worth it. 
Related R-code is available, based on the CRAN packages \textit{multcomp, nparcomp, mratios, emmeans, sandwich, SimLongi, tukeytrend}.
Further extensions can be used, e.g. MCP not just for the first moments of mean value differences but sensitive to more moments by means of most likely transformation models \cite{Kluxen2020}. Next research on this topic will be power (and size) comparisons of standard versus new approaches, particularly for small, unbalanced sample sizes by means of  simulation studies.

\section{Appendix: R-code for data sets}
\par\medskip 
\begin{tcolorbox}[width=15cm]
\sffamily \tiny
\begin{verbatim}
###################### Yap  Resp Res 2021: 22 AITC inhibitors
library(SimComp)
set.seed(170549)
d1<-ermvnorm(4, c(1.1,1.9,2.2), c(0.05, 0.15, 0.35))
d2<-ermvnorm(4, c(1.05,1.25,1.35), c(0.025, 0.25, 0.45))
d3<-ermvnorm(4, c(1.05,0.8,0.75), c(0.025, 0.15, 0.25))
#d4<-ermvnorm(4, c(1.05,1.8,1.65), c(0.025, 0.35, 0.5))
d4<-ermvnorm(4, c(1.05,1.8,1.65), c(0.075, 0.45, 0.5))
d5<-ermvnorm(4, c(1.15,0.5,0.8), c(0.025, 0.05, 0.3))
d6<-ermvnorm(4, c(1.0,0.7,0.5), c(0.025, 0.2, 0.15))
val<-cbind(d1,d2,d3,d4,d5,d6)
co<-val[, c(1,4,7,10,13,16)]
n1<-val[, c(2,5,8,11,14,17)]
n5<-val[, c(3,6,9,12,15,18)]
mrc5<-c(as.vector(co),as.vector(n1), as.vector(n5))
conc<-factor(rep(c("Co",  "n1", "n2"),c(rep(24,3))))
Conc<-as.numeric(rep(c(0,1,5),c(rep(24,3))))
treat<-factor(rep(c("al",  "la", "ti", "pe", "mm", "tr"),each=4))
iatc<-data.frame(mrc5,conc,treat, Conc)	

######################### proportions safety ansofaxine nausea Tab.3
na <-c(rep(0,46), rep(1,3), rep(0,42), rep(1,10),
          rep(0,42),rep(1,10),rep(0,40),rep(1,11),
          rep(0,37),rep(1,14))
dose <-c(rep("0", 49), rep("D40", 52),rep("D80", 52),rep("D120", 51),
         rep("D160", 51))
naus <-data.frame(na,dose)
naus$Dose<-as.factor(naus$dose)									
\end{verbatim}
\rmfamily
\end{tcolorbox}
\par\medskip 

\footnotesize
\bibliographystyle{plain}


\begin{thebibliography}{10}

\bibitem{cramer2016hidden}
AOJ. Cramer, D. van Ravenzwaaij, D. Matzke, H. Steingroever,
  R. Wetzels, R.PPP Grasman, L.J. Waldorp, and E-J. Wagenmakers.
\newblock Hidden multiplicity in exploratory multiway anova: Prevalence and
  remedies.
\newblock {\em Psychonomic Bulletin \& Review}, 23(2):640--647, 2016.

\bibitem{Dunnett1955}
C.~W. Dunnett.
\newblock A multiple comparison procedure for comparing several treatments with
  a control.
\newblock {\em Journal of the American Statistical Association},
  50(272):1096--1121, 1955.

\bibitem{Gabriel1969}
K.~R. Gabriel.
\newblock Simultaneous test procedures - some theory of multiple comparisons.
\newblock {\em Annals of Mathematical Statistics}, 40(1):224, 1969.

\bibitem{greenland1995}
S. Greenland.
\newblock Avoiding power loss associated with categorization and ordinal scores
  in dose-response and trend analysis.
\newblock {\em Epidemiology}, 450--454, 1995.

\bibitem{HRSetal}

L.A. Hothorn, C. Ritz, F. Schaarschmidt, T. Hothorn and S.M. Jensen
\newblock  Simultaneous inference using multiple marginal models- a
R-based tutorial
\newblock {\em in preparation}, 2023.

\bibitem{Hasler2008}
M.~Hasler and L.~A. Hothorn.
\newblock Multiple contrast tests in the presence of heteroscedasticity.
\newblock {\em Biometrical Journal}, 50(5):793--800, October 2008.

\bibitem{Herberich2010}
E.~Herberich, J.~Sikorski, and T.~Hothorn.
\newblock A robust procedure for comparing multiple means under
  heteroscedasticity in unbalanced designs.
\newblock {\em PLOS One}, 5(3):e9788, March 2010.

\bibitem{LH2021y}
L.A. Hothorn.
\newblock Closed test procedures for the comparison of dose groups against a
  negative control group or placebo.
\newblock {\em ArXiv2012.15093 (2021)}, 2021.

\bibitem{LH2022t}
L.A. Hothorn.
\newblock Hidden multiplicity in the analysis of variance (anova):multiple
  contrast tests as an alternative.
\newblock {\em bioRxiv https://doi.org/10.1101/2022.01.15.476452}, 2022.

\bibitem{LH2016}
L.A. Hothorn.
\newblock {The two-step approach-a significant ANOVA F-test before Dunnett's
  comparisons against a control-is not recommended}.
\newblock {\em {Communications in Statistics- Theory and methods}},
  {45}({11}):{3332--3343}, {2016}.

\bibitem{hothorn2022x}
L.A. Hothorn.
\newblock Simultaneous confidence intervals for the interpretation of primary
  and secondary effects in factorial designs without a pre-test on interaction.
\newblock {\em arXiv preprint arXiv:2204.08336}, 2022.

\bibitem{hkh2019}
L.A. Hothorn, F.M. Kluxen, and M. Hasler.
\newblock Pseudo-data generation allows the statistical re-evaluation of
  toxicological bioassays based on summary statistics.
\newblock {\em bioRxiv}, page 810408, 2019.

\bibitem{Kluxen2020}
F.~M. Kluxen and L.~A. Hothorn.
\newblock Alternatives to statistical decision trees in regulatory
  (eco-)toxicological bioassays.
\newblock {\em Archives of Toxicology}, 2020.

\bibitem{Konietschke2013}
F.~Konietschke, S.~Bosiger, E.~Brunner, and L.~A. Hothorn.
\newblock Are multiple contrast tests superior to the anova?
\newblock {\em International Journal of Biostatistics}, 9(1):63--73, May 2013.

\bibitem{Konietschke2012}
F.~Konietschke and L.~A. Hothorn.
\newblock Rank-based multiple test procedures and simultaneous confidence
  intervals.
\newblock {\em Electronic Journal of Statistics}, 6:738--759, 2012.

\bibitem{ksh2015}
F. Konietschke, M. Placzek, F. Schaarschmidt, and L.A. Hothorn.
\newblock {nparcomp: An R Software Package for Nonparametric Multiple
  Comparisons and Simultaneous Confidence Intervals}.
\newblock {\em {Journal of Statistical Software}}, {64}({9}), {MAR} {2015}.

\bibitem{mi2022}
Weifeng Mi, Fude Yang, Huafang Li, Xiufeng Xu, Lehua Li, Qingrong Tan, Guoqiang
  Wang, Kerang Zhang, Feng Tian, Jiong Luo, et~al.
\newblock Efficacy, safety, and tolerability of ansofaxine (ly03005)
  extended-release tablet for major depressive disorder: A randomized,
  double-blind, placebo-controlled, dose-finding, phase 2 clinical trial.
\newblock {\em International Journal of Neuropsychopharmacology},
  25(3):252--260, 2022.

\bibitem{Pallmann2016}
P.~Pallmann and L.~A. Hothorn.
\newblock Analysis of means: a generalized approach using r.
\newblock {\em Journal of Applied Statistics}, 43(8):1541--1560, June 2016.

\bibitem{Pallmann2017}
P.~Pallmann, M.~Pretorius, and C.~Ritz.
\newblock Simultaneous comparisons of treatments at multiple time points:
  Combined marginal models versus joint modeling.
\newblock {\em Statistical Methods in Medical Research}, 26(6):2633--2648,
  December 2017.

\bibitem{Pallmann2018}
P.~Pallmann, C.~Ritz, and L.~A. Hothorn.
\newblock Simultaneous small-sample comparisons in longitudinal or
  multi-endpoint trials using multiple marginal models.
\newblock {\em Statistics in Medicine}, 37(9):1562--1576, April 2018.

\bibitem{Pipper2012}
C.~B. Pipper, C.~Ritz, and H.~Bisgaard.
\newblock A versatile method for confirmatory evaluation of the effects of a
  covariate in multiple models.
\newblock {\em Journal of the Royal Statistical Society Series C-Applied
  Statistics}, 61:315--326, 2012.

\bibitem{Schaarschmidt2021}
F.~Schaarschmidt, C.~Ritz, and L.A. Hothorn.
\newblock The Tukey trend test: Multiplicity adjustment using multiple marginal
  models.
\newblock {\em Biometrics}, 2021 (1-9. DOI: 10.1111/biom.13442)

\bibitem{Tukey1985}
J.~W. Tukey, J.~L. Ciminera, and J.~F. Heyse.
\newblock Testing the statistical certainty of a response to increasing doses
  of a drug.
\newblock {\em Biometrics}, 41(1):295--301, 1985.

\bibitem{Williams1971}
D.~A. Williams.
\newblock Test for differences between treatment means when several dose levels
  are compared with a zero dose control.
\newblock {\em Biometrics}, 27(1):103--, 1971.

\bibitem{yap2021}
Jennifer Maries~Go Yap, Takashi Ueda, Yoshihiro Kanemitsu, Norihisa Takeda,
  Kensuke Fukumitsu, Satoshi Fukuda, Takehiro Uemura, Tomoko Tajiri, Hirotsugu
  Ohkubo, Ken Maeno, et~al.
\newblock Aitc inhibits fibroblast-myofibroblast transition via
  trpa1-independent mapk and nrf2/ho-1 pathways and reverses corticosteroids
  insensitivity in human lung fibroblasts.
\newblock {\em Respiratory Research}, 22(1):1--12, 2021.

\bibitem{Zeileis2006}
A.~Zeileis.
\newblock Object-oriented computation of sandwich estimators.
\newblock {\em Journal of Statistical Software}, 16:1--16, 2006.

\end{thebibliography}

\scriptsize

\textbf{Acknowledgment}: This summary work was created through many years of cooperation with colleagues F. Schaarschmidt (Hannover), Ch. Ritz (Copenhagen), P. Pallmann (Cardiff), F.M. Kluxen (Cologne) and M. Hasler (Kiel) - for which many thanks.

\end{document}